\documentclass[12pt,epsf]{article}
\usepackage{epsfig}

\newcommand{\setfigure}[2]{\begin{figure}[htbp]
\begin{center}\leavevmode\epsfxsize=5in\epsfbox{#1.eps}\end{center}\caption{#2\label{#1}}
\end{figure}}
\newcommand{\twofigures}[3]{\begin{figure}[htdp]
\centering \leavevmode\epsfxsize=2.5in\epsfbox{#1.eps}
\leavevmode\epsfxsize=2.5in\epsfbox{#2.eps} 
\caption{{
#3}\label{#1}}
\end{figure}}

\renewcommand{\thanks}[1]{\footnote{#1}} 

\newcommand{\be}{\begin{equation}}
\newcommand{\ee}{\end{equation}}
\newcommand{\bea}{\begin{eqnarray}}
\newcommand{\eea}{\end{eqnarray}}

\begin{document}

\pagestyle{empty}

\bigskip\bigskip
\begin{center}
{\bf \large Global Structure of a Multi-Fluid Cosmology}
\end{center}

\begin{center}
James Lindesay\footnote{e-mail address, jlslac@slac.stanford.edu} \\
Computational Physics Laboratory \\
Howard University,
Washington, D.C. 20059 
\end{center}
\bigskip

\begin{center}
{\bf Abstract}
\end{center}
Fluid cosmologies are consistent with the generally accepted
observational evidence during intermediate
and late times, and they need not have singular
behavior in primordial times. 
A general form for fluid cosmology consistent with Einstein's
equation is demonstrated, and a dynamic metric
that incorporates fluid scale is developed. 
The large scale causal structure of a multi-fluid
cosmology exemplary of standard cosmology is then examined. 
This is done through developing coupled rate equations
for radiation, dust, and dark components. 
The beginning of the dissolution of the
primordial fluid into the other components
is singularity-free, since the fluid provides
a non-vanishing scale for the cosmology. 
Penrose diagrams are developed for
cosmologies both with and without
a final state dark energy density.

\bigskip \bigskip \bigskip

\setcounter{equation}{0}
\section{Introduction}
\indent

In the most generally accepted form of
standard cosmology, the universe transitions from some
earliest primordial state into a hot, radiation dominated
epoch, then through a residual dust dominated epoch
which eventually yields to domination
by a remnant dark energy density\cite{BigBang}.
The equations that govern a spatially flat expansion satisfy spatial
scale invariance, but not temporal scale invariance, due to the behavior
of the intensive energy densities that drive the dynamics. 
However, the existence of any horizon due to dark energy
introduces a persistent scale to the macroscopic
cosmological dynamics.

Evidence for the existence of persistent dark energy comes
from several independent observations. 
The luminosities of type Ia
supernovae show that the rate of expansion of the
universe was decelerating in the distant past,
but has been accelerating for about 6 
giga-years\cite{TypeIa}. 
This conclusion is independently supported
by analysis of the Cosmic Microwave Background (CMB)
radiation\cite{WMAP,PDG}.  
Both the standard candle luminosity and 
independent CMB structure
results are in quantitative
agreement with a (positive) cosmological constant fit to the data.
The existence of a persistent dark energy density
that might be described in terms of a cosmological constant
defines a length scale to the large scale structure of the cosmology.   

Whole sky, deep field observations yield additional interesting properties.
The \emph{horizon problem}
examines the paradox of the observed large scale
homogeneity and isotropy of the macro-physical properties
of the universe beyond regions of causal influence. 
Uniformity across the whole sky of the temperature of and
angular correlations of the fluctuations in the CMB
have been accurately measured by several experiments\cite{WMAP}. 
These correlations provides evidence for
space-like coherent phase associations amongst the
cosmological fluctuations reflected in the CMB anisotropies.

Quantum fluids exhibit some behaviors similar to
those previously mentioned.  For example,
liquid $^4$He undergoes a cessation of boiling as it is
cooled through its ``lambda" point and develops
a non-vanishing superfluid component.  This is due
to complete temperature homogeneity maintained by
the macroscopic quantum component\cite{Superfluids}. 
Therefore, the dissolution of a primordial quantum
fluid should exhibit paradoxical signs of its space-like
coherent properties.  Since a primordial fluid might have
a finite length (density) scale, the cosmological scale need
not \emph{singularly} vanish at the beginning of dissolution.

It is expected that during the earliest of epochs, the quantum coherence of
gravitating subsystems should qualitatively altered the dynamics
of the cosmology. 
The entangled nature of co-gravitating
quantum states with space-like separations should
manifest as some form of spatial coherence in the
large scale structure of the geometrodynamics.
It is therefore of interest to examine geometric constraints
on the causal relationships in a multi-component universe. 
This paper will explore the large-scale structure
of cosmologies consisting of interrelated fluid components
that evolve exemplary of what is expected from
standard cosmology.
A cosmological model with multiple dynamic scales, one of which
replaces an apparent cosmological ``constant", is shown
to reproduce standard cosmology during intermediate
times, while making the exploration of the early and late time
dynamics more accessible.

\setcounter{equation}{0}
\section{Fluid Cosmology \label{Section2}}
\indent

\subsection{Dynamic space-time description}
\indent

The inclusion of a cosmological constant into Einstein's
equation
\be
G_{\mu \nu} \equiv \mathbf{R}_{\mu \nu} -
{1 \over 2} g_{\mu \nu} \mathbf{R} = - \left (
{8 \pi G_N \over c^4} T_{\mu \nu} + \Lambda g_{\mu \nu}
\right ).
\label{EinsteinEqn}
\ee
provides a convenient parameterization for the
phenomena described using \emph{dark energy}
as a substantial constituent of the energy content
of the universe.  However, if the geometric and dynamic
conservation principles are strictly valid, the constant
$\Lambda$ cannot have evolved from a primordial parameter,
or be evolving towards a remnant value.
For the present discussion, the
cosmology will be assumed to evolve in the absence of any
true cosmological constant $\Lambda_{true}=0$. 
Big bang cosmology is consistent with that of a gravitating
fluid that is homogeneous and isotropic on large scales. 
For an ideal fluid (one with negligible dissipation),
the energy-momentum tensor
generally takes the form
\be
T_{\mu \nu} = P \: g_{\mu \nu} + (\rho + P) u_\mu u_\nu,
\label{Tmunu}
\ee
where the four velocity of the fluid satisfies the consistency
condition
\be
u_\mu g^{\mu \nu} u_\nu = -1.
\label{consistency}
\ee
By taking the trace of the energy-momentum tensor,
$g^{\mu \nu} T_{\mu \nu}  \equiv T_\mu ^\mu = 3 P - \rho$.
one obtains forms of the pressure  and density in terms of
geometric quantities:
\be
\begin{array}{l}
P=T_\phi ^\phi= T_\theta ^\theta = -{c^4 \over 8 \pi G_N} G_\theta ^\theta , \\
\rho=3 P + {c^4 \over 8 \pi G_N} G_\mu ^\mu .
\end{array}
\label{Einsteindensity}
\ee
The components of the flow fields $u^\beta$ can likewise
be determined from the form of the Einstein tensor\cite{FRWdS}.

Dynamic geometries have been shown to
manifest qualitatively different behaviors
from their corresponding static
analogies\cite{JLMay07,BABJL1}.  Rotating systems
(systems with angular dynamics) exhibit off-diagonal
temporal-angular terms
in the metric describing those systems.  Likewise,
radially dynamic black holes described using
metrics with an off-diagonal
temporal-radial term do not introduce physical singularities
reflecting coordinate anomalies at the horizon\cite{BHQG}. 
Using the same reasoning, one is lead to
consider the following hybrid metric,
motivated by the so-called ``river model" of various
static space-time geometries\cite{rivermodel,JLNSBP08}:
\be
\begin{array}{c}
ds^2 = - \left( 1 - {R_{th} ^2(ct) \, r^2 \over R_c ^2 (ct) }  \right )c^2 dt ^2 
- 2 {R_{th}^2(ct) \, r \over R_c  (ct) } \, c dt  \,  dr \\
+ R_{th}^2 (ct) \left ( dr^2 + r^2 d \theta ^2 + r^2 sin^2 \theta d\phi ^2   \right )  \\ 
= -  c^2 dt^2 + R_{th}^2(ct) \left( dr - {r \over R_c (ct)} c dt  \right )^2  \\
+ R_{th}^2 (ct) \left (  r^2 d \theta ^2 + r^2 sin^2 \theta d\phi ^2   \right ) .
\end{array}
\label{FRWdSmetric}
\ee
This metric was developed to incorporate scales that
diagonalize towards a Robertson-Walker
form\cite{RWgeometry} when
maintaining the temporal coordinate, and towards
a de Sitter form when maintaining the
radial coordinate\cite{dualdiag}. 
A de Sitter geometry\cite{deSittergeometry},
which manifests a horizon, is of interest for describing
a cosmology with a positive cosmological constant. 
For the metric given in Eq. \ref{FRWdSmetric}, the
temporal coordinate is the time of an observer located
at the coordinate ``center" $r=0$.  The \emph{cosmological
principle} asserts that this center and observer is not unique or
special, and that spatial coordinates exist
that can be freely translated
and rotated.  For this observer, the function $R_{th}(ct)$
represents a scale for proper length measurements. 
The function $R_c$ represents the scale in a de Sitter
space-time in the static limit.  For a de Sitter geometry
($R_{th}=1$), the de Sitter scale is the radial coordinate
of the horizon surrounding the observer.

The hydrodynamic parameters can be immediately calculated using
Eq. \ref{Einsteindensity}:
\be
\begin{array}{l}
\rho= {3 c^4 \over 8 \pi G_N} \left (
{1 \over R_c}  + {\dot{R}_{th} \over R_{th} }    \right )^2 , \\ \\
P = -\rho - {c^4 \over 4 \pi G_N} {d \over dct} \left (
{1 \over R_c} + {\dot{R}_{th} \over R_{th} }  \right ) , 
\end{array}
\label{FRWdensity}
\ee
where $u_0=-1$, and the other components $u_j$ vanish.
The identifications in Eq. \ref{FRWdensity} then allows
the dynamics to be expressed solely in terms of the energy
content:
\be
{d \over d ct} \rho = -\sqrt{{24 \pi G_N \rho  \over c^4 }} (P + \rho) .
\label{rhodynamics}
\ee
This relationship describes the expected dynamics of
standard cosmology (as well as its
various fractions\cite{ThermalDual}),
once the appropriate equation of
state relating the pressure to the density is incorporated.

\subsection{Diagonal metric form}
\indent

The metric form in Eq. \ref{FRWdSmetric} can be
directly diagonalized using a coordinate transformation
of the radial coordinate $r$ while maintaining the
temporal coordinate $ct$.  Examining
Eq. \ref{FRWdensity}, one can define the
Robertson-Walker scale factor $a(ct)$, and require angular
isotropy of the metric expressed in either coordinate
system:
\be
{\dot{a} \over a} \equiv {\dot{R}_{th} \over R_{th}}
+ {1 \over R_c} \quad , \quad
R_{th} \, r =a \, r_{RW} \quad .
\label{RWscale}
\ee
A brief and straightforward calculation yields
the form of the coordinate transformation,
\be
R_{th} \, dr = a \left [ {r_{RW} \over R_c} dct +
dr_{RW}    \right ] \, .
\ee
Substitution of this form into Eq. \ref{FRWdSmetric} gives
the expected metric
\be
ds^2 = -c^2 dt^2 + a^2(ct)  \left [ dr_{RW} ^2 +
 r_{RW}^2 d \theta ^2 + r_{RW} ^2 sin^2 \theta \, d\phi ^2   \right ].
\label{RWmetric}
\ee
The diagonal form obtained when the radial coordinate
is maintained has been explored elsewhere\cite{dualdiag}.
The metric form in Eq. \ref{RWmetric} explicitly demonstrates that
the temporal parameter $t$ being used to describe the
dynamics in Eq. \ref{FRWdSmetric} is the same as that
used in the Robertson-Walker metric, whose coordinates
manifest spatial homogeneity and isotropy.

\setcounter{equation}{0}
\section{Multi-Fluid Cosmology \label{Section3}}
\indent

The generic form Eq. \ref{RWscale} will next be specialized
in a manner that conveniently incorporates fluid scale
into the cosmology.  The radial rescale factor $R_{th}$ will
be taken to have the constant value of unity, providing a
direct relationship between the radial coordinate $r$ and
that of the Robertson-Walker form,
$R_{th}=1,  r_{RW}={r \over a(ct)}$.  The relation
\ref{FRWdensity} then directly connects the
dynamic scale $R_c (ct)$ to the fluid density
$R_c \equiv R_\rho$ via
\be
\rho= 
{3 c^4 \over 8 \pi G_N} {1 \over R_\rho ^2 (ct)}
\quad , \quad 
{1 \over R_\rho}= {\dot{a} \over a}.
\label{fluidscale}
\ee
Thus, the fluid scale is directly represented in the
metric form
\be
ds^2 = - \left( 1 - { r^2 \over R_\rho ^2 (ct) }  \right )c^2 dt ^2 
- 2 { r \over R_\rho  (ct) } \, c dt  \,  dr 
+ dr^2 + r^2 d \theta ^2 + r^2 sin^2 \theta \, d\phi ^2    .
\label{fluidmetric}
\ee

As previously stated,
standard big bang cosmology models the evolution
of a universe containing a remnant pressureless
\emph{dust} content (to
which most familiar matter contributes, along with
any \emph{dark matter}), \emph{radiation}
(reflected in later stages via the thermal cosmic microwave
background), and a now substantial
component of \emph{dark energy}:
\be
\rho=\rho_{DE}+\rho_{rad}+\rho_{dust},
\ee
where for the present, primordial and remnant dark
energies are included in the term $\rho_{DE} = \rho_{primordial}
+\rho_{remnant}$.

It is convenient to develop coupled rate equations
to describe the evolution of constituent components
that combine to give Eq. \ref{rhodynamics}.  
A general form for a
component rate equation is given by
\be
{d \rho_{DE} \over dct}=-D_{DE \rightarrow rad}(ct)
-{3 \over R_\rho} (P_{DE}+ \rho_{DE}) .
\label{darkevolution}
\ee 
The
radiation will be assumed to ``precipitate"
from the primordial dark component in early times. 
The first term on the right of
Eq. \ref{darkevolution} represented by
$D_{DE \rightarrow rad}$ is a generic
rate of the dissolution
of the primordial dark
energy into radiation and remnant dark energy,
whose detailed form is determined by related
micro-physical processes.  The second term
generally incorporates the equation of state of
the given component in a manner consistent
with the composite rate equation.

Radiation
has its energy component red-shifted during an
expansion.  Since the inverse fluid scale $1/R_\rho$
is the logarithmic derivative of the Robertson-Walker (RW)
metric scale $a$, and radiation density scales with
the inverse $4^{th}$ power of the RW scale, the
rate equation describing the radiation is expected
to take the form
\be
{d \rho_{rad} \over dct}=D_{DE \rightarrow rad}(ct)
-{4 \over R_\rho} \rho_{rad}
- \Theta (\rho_{rad}-\rho_{threshold}) 
{\rho_{rad} \over c \tau_{r \rightarrow d}} .
\label{radevolution}
\ee
The first term on the right of Eq. \ref{radevolution} incorporates
the dissolution of primordial dark energy into radiation, the
second incorporates the appropriate red shift
(and equation of state) of radiation,
and the third term assumes that above a threshold density,
microscopic asymmetries or other
processes generate remnant dust from the
radiation.  Since the nature of the transition of the primordial density
into any dark matter component in the dust is not understood,
for present purposes its dissolution is
completely incorporated into the radiation equation. 
However, any partitioning of this dissolution into
radiation and dark matter should not affect the global
structure of the cosmology being explored.

The energy components of the constituents of pressureless
dust are not expected to red shift during the
expansion.  Therefore, the dust density scales with
the inverse $3^{rd}$ power of $a$, yielding a rate equation
of the form
\be
{d \rho_{dust} \over dct}=\Theta (\rho_{rad}-\rho_{threshold}) 
{\rho_{rad} \over c \tau_{r \rightarrow d}}
-{3 \over R_\rho} \rho_{dust} .
\label{dustevolution}
\ee
The first term on the right of Eq. \ref{dustevolution} generates
the remnant dust in early times, while the second term
insures proper scaling of this density during expansion.

The forms of Eqs. \ref{darkevolution},
\ref{radevolution}, and \ref{dustevolution}
insure that the evolution of the total density given by
Eq. \ref{rhodynamics} is consistent with the summed density
components, as long as the pressure content is appropriate. 
The radiation component has been taken to satisfy
$P_{rad}={1 \over 3} \rho_{rad}$, and the dust has been assumed
not to contribute to the pressure.  The primodial form of the
dark pressure will likely be
that of a macroscopically coherent system undergoing the
phase transition to a hot dense plasma.  The equation
of state for the dark energy gets consistently incorporated in the
form of the rate equation for the dark energy component
$\dot{\rho}_{DE}$,
as has been done for the radiation and dust
components.  For the popular
model of an early state
\emph{inflation} with a form that would have generated a
de Sitter space-time had it persisted, the primordial dark
energy might be taken to satisfy the same equation of state
that the remnant dark energy
seems to satisfy, $P_{DE}=-\rho_{DE}$.
In that case, the form of the pressure can be taken as
\be
P=P_{dust}+P_{rad}+P_{DE}
={1 \over 3} \rho_{rad} - \rho_{DE}.
\label{pressure}
\ee
However, quite generally, once the coupled rate
equations \ref{darkevolution}, \ref{radevolution},
and \ref{dustevolution} have been developed,
they can be solved to determine the fluid scale
$R_\rho (ct)$, from which the large scale
structure of the cosmology can be explored.

\setcounter{equation}{0}
\section{Penrose Diagrams for Fluid Cosmologies
\label{Section4}}
\indent

A Penrose diagram is a convenient tool for examining
the large-scale causal structure of a given
space-time\cite{LSJLBlackHoles}. 
Penrose diagrams are space-time
diagrams with the following properties:
\begin{list}{...}{}
\item the coordinates are conformal,
which insure that outgoing light-like
surfaces have a slope of +1, and ingoing light-like
surfaces have a slope of -1,
\item the domains of the coordinates are finite, which
allows the whole geometry to be displayed in a finite
image.
\end{list}
For isotropic systems, the angular coordinates
$(\theta, \phi)$ are given arbitrary fixed values,
from which one infers
that any point on the remaining 2-dimensional
diagram represents a spherical surface
at a given time. 
This means that time-like relationships are
vertical relative to the diagonal light-like curves,
while space-like relationships are horizontal.  Causal
relationships always have relatively
vertical or light-like orientations on the diagram. 
Likewise, regions that might contain systems
with space-like coherence can be directly identified
on such a diagram.  

A set of conformal coordinates
that might be used to construct
the Penrose conformal coordinates can be directly
obtained from the RW metric form Eq. \ref{RWmetric} 
\be
ds^2 = a^2(ct)  \left [-{c^2 dt^2 \over a^2(ct)} +  dr_{RW} ^2 +
 r_{RW}^2 d \theta ^2 + r_{RW} ^2 sin^2 \theta \, d\phi ^2   \right ].
\label{ConformalMetric}
\ee
The first term in the bracket of Eq. \ref{ConformalMetric}
is the differential form of the conformal time $dct_*$. 
This then yields a metric whose null radial geodesics
have slope $\pm 1$ using differential coordinates $(dct_*,dr_*)=
(dct/a(ct), dr_{RW})$.
The Robertson-Walker scale factor $a$
can be directly calculated once the
fluid scale is known by using Eq. \ref{fluidscale}
\be
a(ct)=R_\rho(0) \, exp \left [
\int _0 ^{ct} {dct' \over R_\rho(ct')} \right ] .
\ee
This gives a generic form for conformal coordinates
centered at fluid coordinate $(ct_o,r_o)$:
\be
\begin{array}{l}
ct_* = \int_{ct_o} ^{ct} 
 {dct' \over a(ct')}, \\
r_* = {r \over a(ct)} -  {r_o \over a(ct_o)} .
\end{array}
\label{conformal}
\ee
The Penrose conformal coordinates will then be constructed
using
\be
\begin{array}{l}
Y_\rightarrow =[tanh({ct_*+r_* \over scale})-
tanh({ct_*-r_* \over scale})]/2 \, , \\ \\
Y_\uparrow =[tanh({ct_*+r_* \over scale})+
tanh({ct_*-r_* \over scale})]/2 \, .
\end{array}
\label{PenroseCoords}
\ee
In what follows in this section, these coordinates
will be used to construct Penrose diagrams
for cosmologies that evolve from a primordial
``dark" fluid, through radiation and dust domination,
and perhaps towards a remnant dark energy dominated
geometry.

\subsection{A cosmology with no remnant dark energy
\label{NoDE}}
\indent

At any given time,
the \emph{particle horizon} $r_{PH} (ct)$  for an
observer located at $r=0$ is the radial location on the
space-like curve $t=0$ whose ingoing light-like
trajectory intersects the observer at that time. 
This particular surface on the past light cone
represents the outermost region of the universe
from which communications from the beginning
can have been received by this observer. 
On a Penrose diagram, this location can be directly
determined by extending an ingoing light-like trajectory
(line of slope -1) from the time in question for the
observer $(ct, r=0)$ back to the time of
dissolution $(0, r_{PH} (ct))$.
There is also a surface corresponding to
the set of observers for which the center
$r=0$ is contained on their particle horizons,
which will be referred to as the \emph{particle
out horizon}.  This outgoing light-like surface begins
at the origin $(ct=0,r=0)$ and terminates on future
infinity.  Again, by the cosmological principle, none
of the observers is spatially special or unique.

For the purposes of the constructions, the primordial
dark fluid was assumed to ``decohere" into radiation
using a form $\rho_{DE}(ct) \sim exp[-(t/ \tau_{DE
\rightarrow rad})^2]$, and the initial Robertson-Walker
scale has been chosen to be the same as the
initial fluid scale $a(0)=R_\rho (0)$.  In this subsection
it will be assumed that there is no persistent dark
energy, i.e., $\rho_{DE}(\infty )=0$.  Some of the
features of such a fluid cosmology are displayed
on the Penrose diagram in Figure \ref{FeatNoDE}. 
The origin of the conformal coordinates in
Eq. \ref{conformal} is chosen to coincide with that
of the coordinates in the metric Eq. \ref{fluidmetric}. 
\setfigure{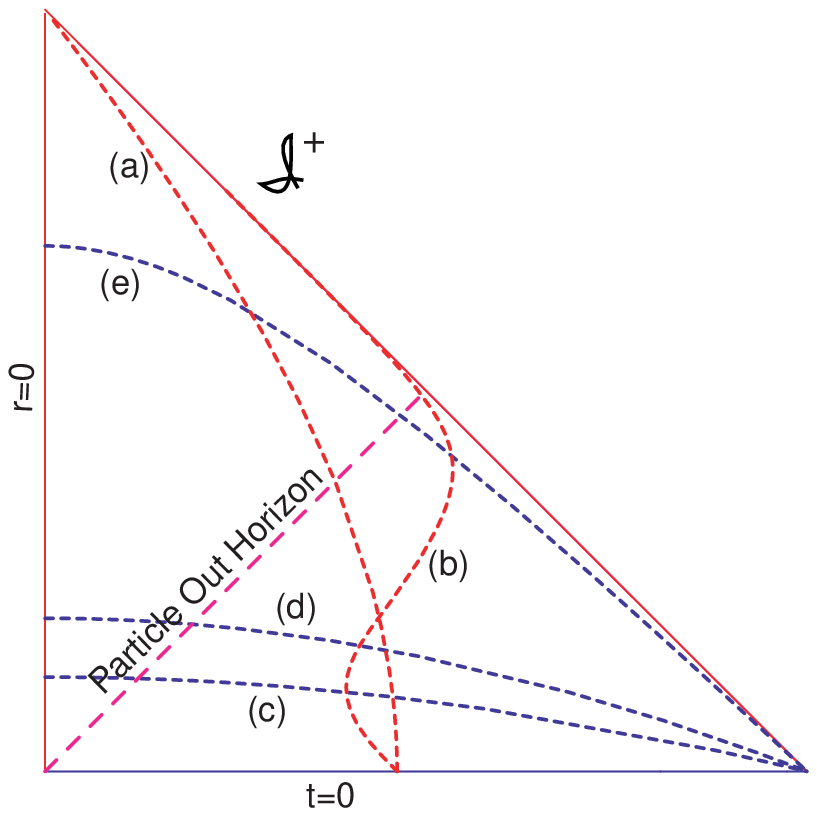}{Penrose diagram of the dynamic
features of a multi-fluid cosmology with no final
state dark energy. 
The vertical line bounding the diagram from the left is
the time-like trajectory of the ``center" $r=0$.  The horizontal
line bounding the diagram from below represents the
space-like beginning of dissolution $t=0$.  The
\emph{particle out horizon}
is an outgoing light-like surface originating at the origin
$(ct=0, r=0)$, terminating on future light-like infinity
labeled $skri^+$.  Trajectory
(a) represents the Robertson-Walker scale $a(ct)$.  Trajectory
(b) represents the fluid density scale $R_\rho (ct)$.  The time (c)
is that of the transition from primordial dark energy to radiation
domination, time (d) that of maximum dust density, and time
(e) that of the transition from radiation domination
 to dust domination.}
Since there is a beginning to the dissolution of the
primordial dark fluid, this time is represented
as the \emph{non-singular} space-like
curve $t=0$ bounding the lower portion of the diagram. 
The time-like trajectory of a stationary (inertial) observer at the
center of this representation is labeled $r=0$.  The remaining
boundary of this diagram is future infinity $skri^+$, which
is an ingoing light-like surface.

The trajectories of several features are displayed on this
diagram.  The radial fluid scale $R_\rho (ct)$ from
Eq. \ref{fluidscale} is represented by the surface labeled
(b). As has been stated, the Robertson-Walker scale $a(ct)$,
labeled (a), is seen to take an initial value coincident with
the fluid scale $R_\rho$, and to terminate at the upper
corner of the diagram.  The space-like volume
labeled (c) is that of the cosmology
as it transitions from primordial dark fluid domination
to radiation domination.  For the parameters chosen
in this calculation, the space-like volume labeled (d) is that of
the cosmology at the time of maximum dust density,
and the space-like volume labeled (e) is that of
the cosmology at the time of its transition from
radiation to dust domination. 
The particle out horizon represents the surface of
earliest causal influence of any constituent of
the cosmology originally
located at $r=0$ upon other constituents in the
cosmology.

Curves of constant temporal and radial components
are superimposed on the previous Penrose diagram
in Figure \ref{dSrdS}. 
\twofigures{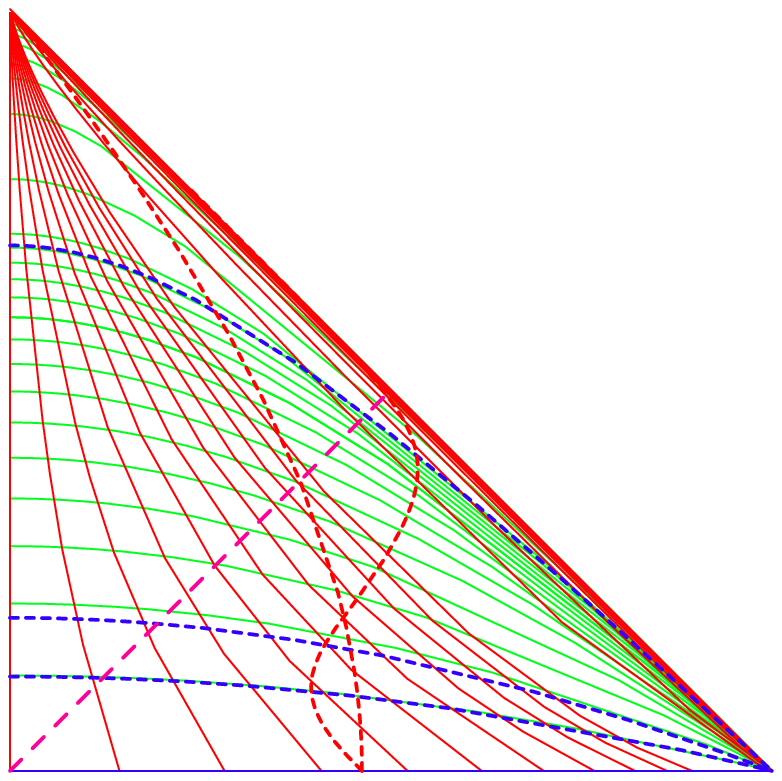}{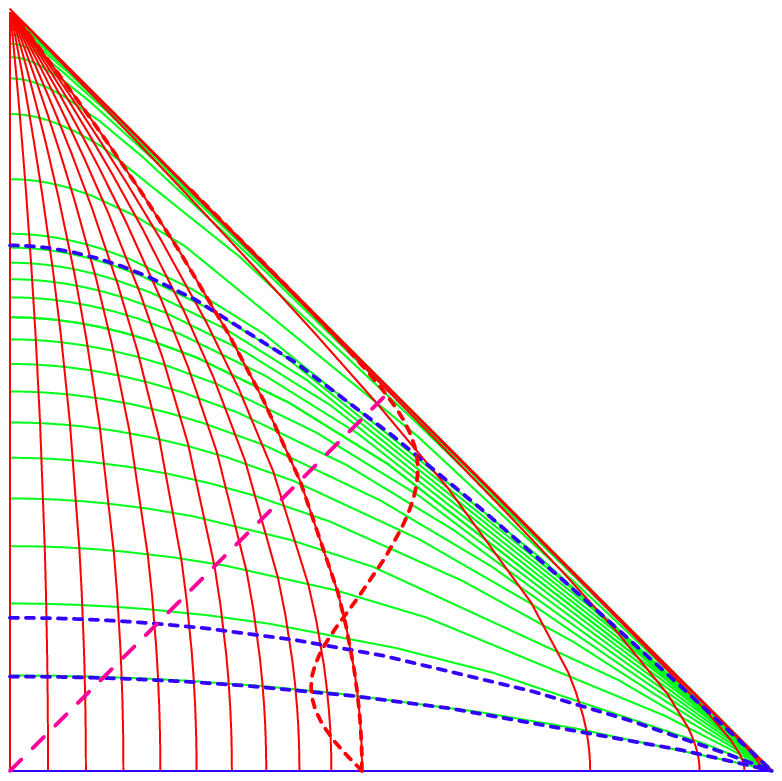}{Radial coordinates for
a cosmology with no remnant dark energy, superimposed
on the features diagram Fig. \ref{FeatNoDE}.  Both Penrose
diagrams show fixed temporal coordinate curves
(green) $ct=constant$ graded in tenths, then in the given units
of the scale of the diagram,
originating on the time-like curve $r=0$, and terminating at the far
right corner of the diagram.  The diagram on the left shows
fixed area radial coordinate curves (red) $r=constant$,
while that on the right shows fixed Robertson-Walker radial
coordinate curves $r_{RW}=constant$, each
graded in tenths, then units of the given scale.  The radial
coordinate curves originate on the space-like curve
$ct=0$ and terminate at the uppermost corner of the diagram.}
The diagram on the left represents a coordinate grid
$(ct,r)$ for the dynamic fluid metric form. 
From Eq. \ref{fluidmetric}, one determines that
any point on the diagram corresponds to
the surface of a sphere of area $4 \pi r^2$.
Volumes of constant $ct$ are represented by the space-like
(green) curves graded in tenths, then in
units of the given scale.  Each
of these curves originate on the observational center
$r=0$ and terminate at the far right corner of the diagram. 
Surfaces of fixed area $4 \pi r^2$ are represented by the
(red) curves originating on the dissolution volume
$t=0$ and terminating at the top corner of the diagram,
initially graded from $r=0$
in tenths, then in units of the given scale.
Future light-like infinity $skri^+$ is seen to
correspond to the late-time /
asymptotic-radial coordinate curve $(ct \rightarrow \infty ,
r \rightarrow \infty )$. 
The particle out horizon is therefore seen to cross all
temporal and radial coordinate
curves at some point in the global cosmology.

The diagram on the right represents a coordinate grid
$(ct,r_{RW})$ for the RW metric form. 
The volumes of constant $ct$ are
again represented by the same space-like
(green) curves as in the diagram on the left.  
From Eq. \ref{RWmetric},
surfaces of area $4 \pi [a(ct) \, r_{RW} ]^2$ are represented by the
(red) curves originating on the dissolution volume
$t=0$ and terminating at the top corner of the diagram,
initially graded from $r_{RW}=0$
in tenths, then in units of the given scale.  The
curve $r_{RW}=1$ is of course seen
to correspond with the trajectory of the
Robertson-Walker scale.
Although in principle it is possible to maintain
an accelerating trajectory that remains external to the
region of causal influence of a constituent originally
located at $(ct=0,r=0=r_{RW})$, there are no inertial
(time-like) observers
that can remain exterior to the particle out horizon
for this cosmology.  Of course, \emph{all} outgoing light-like
communications originating on $(ct=0,r>0)$
reach future light-like infinity without ever
crossing $r=0$.

\subsection{A cosmology with remnant dark energy}
\indent

A cosmology with a non-vanishing final state dark
energy manifests a horizon beyond which there
cannot be any incoming communication to an
inertial observer located at $r=0$.  This horizon
will be an ingoing light-like surface separating
the two regions of causal influence.  For this reason,
unlike the cosmology discussed in subsection \ref{NoDE},
the intersection of the particle out horizon with the horizon
$(ct_X,r_X)$ provides a unique scale for the Penrose diagram
representing this cosmology.  This scale will serve
as the center of the conformal coordinates
in Eq. \ref{conformal} used
to construct the diagram.  
The primordial dark
energy density will be assumed to take a form similar
to that discussed in the previous subsection, only
relaxing towards a final non-vanishing value for
the remnant dark energy density ${\rho_{DE}} 
(ct \rightarrow \infty) \Rightarrow  
 \rho_\Lambda$.  All other parameters were taken
to be the same as for the previous cosmology. 
Some features of this multi-fluid cosmology are
displayed on the Penrose diagram in Figure \ref{FeatDrkE}.
\setfigure{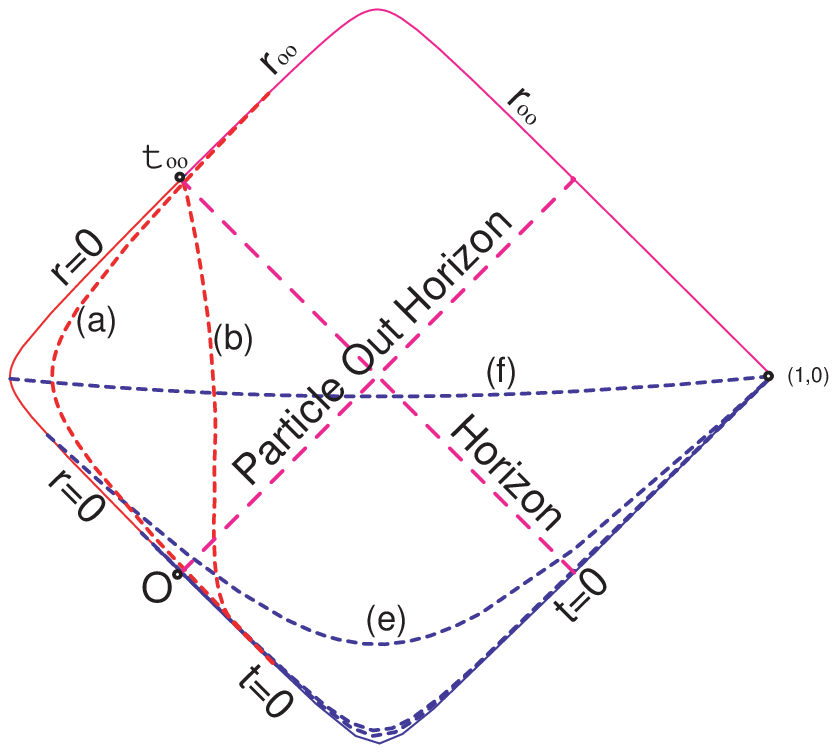}{Penrose diagram of the dynamic
features of a multi-fluid cosmology with non-vanishing final
state dark energy.  The scale determined by the intersection of the
outgoing light-like surface defining the particle out horizon
with the ingoing light-like surface defining the
horizon is set at the center of the conformal
coordinates.  The particle out horizon again begins at
$(ct=0,r=0)$ labeled O, and terminates on the space-like curve
$r_\infty $
bounding the diagram from above.  The horizon originates
at a finite radial coordinate on the curve $ct=0$
and terminates at the future infinity termination
point for fixed area radial curves $t_\infty $.
The time-like trajectory between the initiation of the particle out
horizon at O and future infinity termination
$t_\infty $ bounding the diagram from
the left is the ``center" $r=0$.  The space-like trajectory between
the initiation of the particle out horizon at O and the far right corner
$(1,0)$ bounding the diagram from below is the beginning
of dissolution $t=0$.  Again, the curve (a) represents the
Robertson-Walker scale $a(ct)$, while curve (b) represents
the fluid density scale $R_\rho$.  The times of transition from
primordial dark energy to radiation domination and maximum dust
density are barely distinguishable near the $t=0$ curve, while
again the volume at the time of
 transition from radiation domination to dust domination is
labeled (e). 
There is an additional time of transition from dust
domination to remnant dark energy domination
represented by the mid-time dashed
space-like curve labeled (f).
}
The beginning of the dissolution of the primordial dark
fluid is again represented by the
\emph{non-singular} space-like curve
$t=0$ bounding the diagram from below, originating at
the origin $(ct=0,r=0)$ (the source point O of the particle
out horizon), and terminating at the far
right corner of the diagram, which has the
extremal value for the Penrose conformal
coordinates $(1,0)$.  There are no other coordinates
on the Penrose diagram associated with
the limiting values ($-1 \leftarrow Y \rightarrow +1$)
of the conformal coordinates $(Y_\rightarrow, Y_\uparrow)$,
or light-like surfaces
between these extreme values. 
Because of the exponential temporal behavior of the
scale $a(ct)$ in Eq. \ref{conformal}, \emph{proportionate}
late-time/asymptotic-radial coordinates $(ct \rightarrow
\infty , r \rightarrow \infty )$ correspond to the single
conformal coordinate $(ct_* (\infty ),-r_X/a(ct_X))$, labeled $t_\infty $ on the
diagram.  The time-like trajectory
of a stationary observer at the center of this representation,
originating at the origin  O and terminating
at its future infinity $t_\infty $, is labeled $r=0$.  
The point  $t_\infty $ is unique with regards to coordinates
$(ct,r)$.  The diagram is
bounded from above by the space-like curve $r_\infty$ representing
infinite static areas with radial coordinate $r \rightarrow \infty $. 
This future boundary is space-like despite being represented
by the asymptotic behavior of the radial coordinate. 

There are several other surfaces of interest on this diagram. 
The fluid scale $R_\rho (ct)$ from Eq. \ref{fluidscale} is
again represented by the surface labeled (b), which
originates at a finite radial scale
on the space-like surface $t=0$, and terminates
defining the horizon at the future infinity point for $r=const$
surfaces labeled $t_\infty $. 
The ingoing light-like surface that originates on the dissolution
volume and terminates at this point globally constructs
the horizon of this geometry. 
The Robertson-Walker scale $a(ct)$
labeled (a) on the diagram is a time-like surface that
originates on the dissolution volume $t=0$
at the fluid scale $R_\rho (0)$ and terminates on the static
infinite area surface $r_\infty $.  The space-like
volumes barely distinguishable in the
diagram from that at the time of dissolution $t=0$
represent the time of transition from primordial dark
energy domination to radiation domination and the time
of maximum dust density using the same parameters
as used for constructing the diagrams with no
remnant dark energy density.  The time of transition from
radiation domination to dust domination is labeled (e),
and the time of transition from dust domination to
remnant dark energy domination is labeled (f). 
Again, the particle out horizon is an outgoing
light-like surface that represents the surface of
earliest causal influence of any constituent originally
located at $r=0$ upon other constituents in the
cosmology.  The point of observational origin for
the representation, O, is the point of
initiation for the particle out horizon, which terminates
on the static infinite area surface $r_\infty $.  As
previously mentioned, the scale has been chosen so that
the particle out horizon crosses the horizon at
the center of the conformal coordinates of the diagram
$(Y_{\rightarrow X}=0,Y_{\uparrow X}=0)$.

For this cosmology with a remnant dark energy,
curves of constant temporal and radial components
are superimposed on the previous Penrose diagram
in Figure \ref{DarkErdS}. 
\twofigures{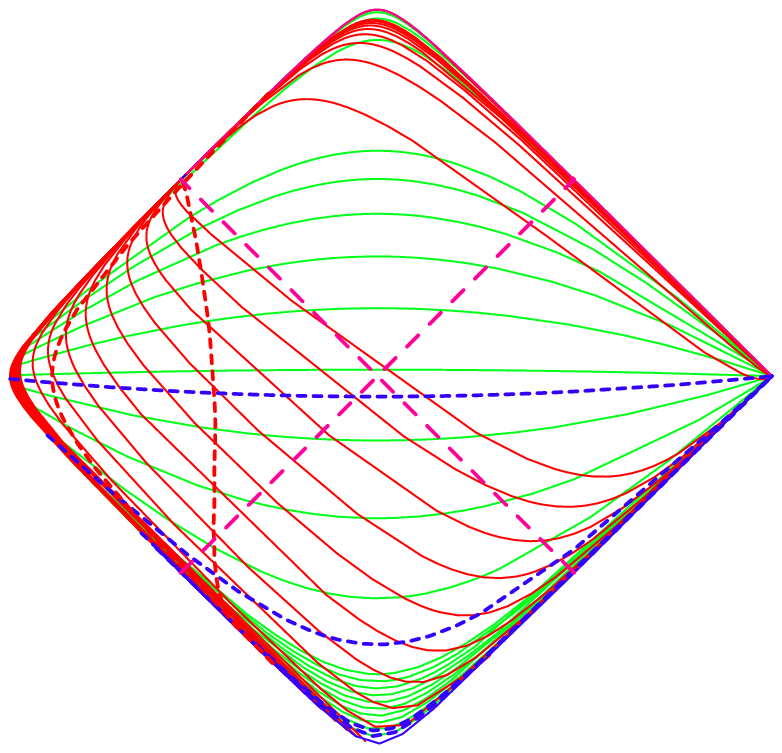}{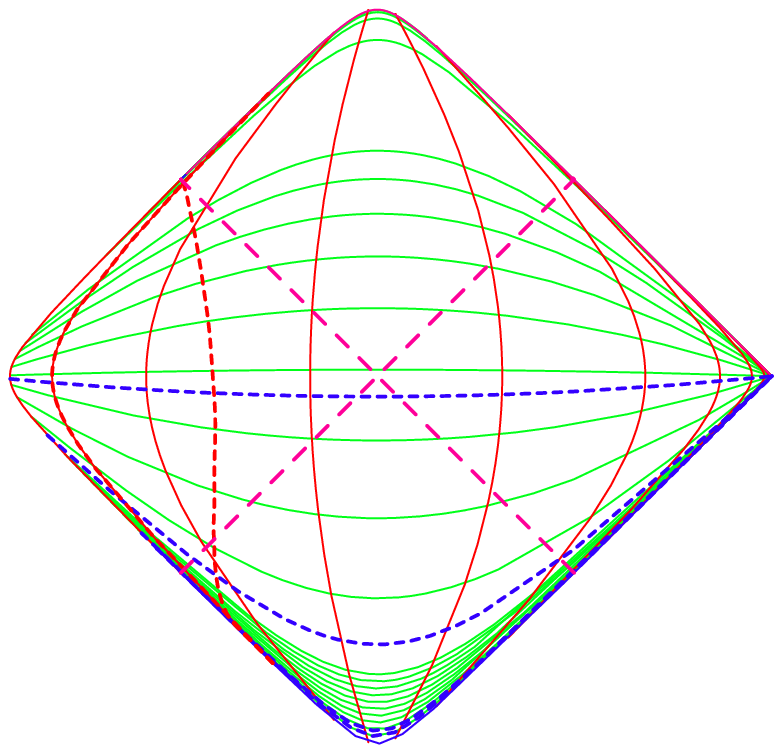}{Radial coordinates for
a cosmology with a remnant dark energy density, superimposed
on the features diagram Fig. \ref{FeatDrkE}.  Both Penrose
diagrams show fixed temporal coordinate curves
(green) $ct=constant$ graded in tenths, units, then
decades of the given scale of the diagram,
originating on the time-like curve $r=0$, and terminating at the far
right corner of the diagram.  The diagram on the left shows
static area coordinate curves (red) $r=constant$,
graded in tenths, units, then decades of the given scale.
The diagram on the right shows fixed Robertson-Walker radial
coordinate curves $r_{RW}=constant$,
graded in units, then decades of the given scale.  The radial
coordinate curves originate on the space-like curve
$ct=0$ and terminate on the future infinity of the
coordinate.}
The diagram on the left represents a
coordinate grid of volumes of fixed times $ct$ and
surfaces of fixed areas
$4 \pi r^2$ labeled $(ct,r)$ using
coordinates from the dynamic fluid metric
form Eq. \ref{fluidmetric}.
Volumes of constant $ct$ are represented by the space-like
(green) curves graded in tenths, units, then decades
of the given scale.  Each
of these curves originates on the observational center
$r=0$ and terminates at the far right corner of the diagram,
which has the extremal value for the Penrose conformal
coordinates $(1,0)$. 
Surfaces of fixed area parameterized by the radial
coordinate $r$ are represented by the
(red) curves originating on the dissolution volume
$t=0$ and terminating at the point $t_\infty $
of the termination
of the radial fluid scale $R_\rho(\infty)$,
initially graded from $r=0$
in tenths, then in units and decades of the given scale.
As can be extrapolated from the diagram,
the space-like curve labeled $r_\infty $ bounding future
trajectories is seen to
correspond to the late-time/asymptotic-radial
coordinate curve $(ct \rightarrow \infty ,
r \rightarrow \infty )$.  The trajectory of a radial
coordinate in the metric Eq. \ref{fluidmetric} can be
space-like due to the non-orthogonal nature of the
coordinates in the metric.  Light-like radial trajectories follow null
geodesics of the metric, satisfying
\be
{dr_\gamma \over dct}= {r_\gamma \over R_\rho (ct)} \pm 1,
\ee
for outgoing/ingoing trajectories $r_\gamma (ct)$. 
Therefore, ingoing light-like trajectories are momentarily
stationary in the radial coordinate as they traverse the
fluid scale $R_\rho (ct)$.  This means that each of the
static area, fixed $r$ curves have slope -1 as the
fluid scale crosses that coordinate.  Fixed radial coordinate
curves are seen to be time-like surfaces left of the fluid
scale, and space-like surfaces to the right of that scale.
As was the case with no remnant dark energy,
the particle out horizon is seen in Fig. \ref{DarkErdS}
to cross all temporal and radial coordinate
curves at some point in the global cosmology.
However, for this cosmology there are clearly
\emph{inertial} trajectories (right of the horizon)
external to having causal influence upon
an inertial constituent originally located at $(ct=0,r=0)$.

The diagram on the right represents a coordinate grid
$(ct,r_{RW})$ for the Robertson-Walker metric form. 
The volumes of constant $ct$ are
again represented by the same space-like
(green) curves as in the diagram on the left.  
Surfaces of fixed radial scale $r_{RW}$
(with time dependent areas
$4 \pi [a(ct) \, r_{RW} ]^2$) are represented by the
(red) curves originating on the dissolution volume
$t=0$ and 
terminating on the space-like curve labeled
$r_\infty $,
graded from $r_{RW}=0$
in units, then decades of the given scale.  The
curve $r_{RW}=1$ is again seen
to correspond with the trajectory of the
Robertson-Walker scale.
Unlike fixed area radial coordinates $r$, the
RW fixed radial coordinate curves $r_{RW}$
terminate on differing points on the future
bounding curve $r_\infty $.

Curves with fixed RW coordinate $r_{RW}$ can represent
the trajectories of co-moving (intertial) observers. 
There are clearly inertial observers that can originate
on the volume $t=0$ and terminate external to any
causal influence upon/from an observer at
$r=0$ (e.g. all fixed radial RW coordinates originating
and terminating to the right of the horizons).

\setcounter{equation}{0}
\section{Conclusions}
\indent

The global causal structure of a spatially coherent
dynamic fluid cosmology has been examined in
this paper.  A geometry with early spatial coherence
can directly address the horizon problem
as usually put forth in standard cosmology.
The multi-fluid cosmologies examined evolve
from a singularity-free initial dark fluid
that goes through a dissolution
into radiation and dust, with perhaps a non-vanishing
final dark energy content, while satisfying
all expected dynamic and geometric
conservation principles.

In both cases examined, the Penrose diagrams were found
to be bounded by three surfaces.  For a fluid cosmology with
no remnant dark energy, the bounding surfaces consist of
a time-like ``center", a space-like beginning, and a light-like
future infinity, as was expected. 
However, the Penrose diagram for a fluid cosmology
\emph{with} remnant dark energy had some
unexpected characteristics.  The bounding surfaces consist
of one time-like and two space-like curves.  There
is only one point on the diagram where a Penrose conformal
coordinate reaches an extremal value, which is the
terminating (distant) infinity of the singularity-free
space-like beginning of dissolution.  The future infinity
of time-like observers is a space-like static infinite area
surface.  The time-like ``center" trajectory is an
ordinary inertial world line that traverses
between points in the
beginning and future bounding volumes.
 In this cosmology, the existence of a horizon
provides a natural scale for the diagram.

Within this discussion, no consideration has been given
for quantum measurability
constraints\cite{JLHPNOct04,NSBP06,JLHPNEJ},
thermodynamic or holographic
considerations that might arise from the finite horizon scale
(such as possible Poincare recurrences\cite{LDJLLS}),
or micro-physical specifics. 
Work on quantum behaviors in fluid cosmologies is
presently underway. 
Further questions concerning self-consistent
relationships between co-gravitating quantum components
analogous to those developed for dynamic black holes\cite{BHQG},
 are being examined.  Those relationships should give insight
into constraints upon the temporal behavior
of the primordial fluid during
its dissolution\cite{NSBP06}.  Results of these explorations will
be presented in a subsequent paper.

\bigskip
\begin{center}
\textbf{Acknowledgments}
\end{center}

The author gratefully acknowledges useful past discussions with 
James Bjorken, E.D. Jones, H. Pierre Noyes,  
Michael Peskin, and Lenny Susskind.

\end{document}